\documentstyle[12pt]{article}
\def\lsim{\mathrel{\raise3pt\hbox to 8pt{\raise -6pt\hbox{$\sim$}\hss{$<$}}}}
\def\haf{\textstyle{1\over2}}
\def\forth{\textstyle{1\over4}}
\def\A{\rm A}
\def\minus{\mbox{$-$}}
\newcommand{\vp}{\vec{\, p}}
\newcommand{\vr}{\vec{\! r}}

\newcommand{\vD}{\vec{D}}
\newcommand{\vJ}{\vec{J}}

\newcommand{\vS}{\vec{S}}
\newcommand{\vsig}{\vec{\sigma}}
\newcommand{\btau}{\mbox{\boldmath $\tau$}}
\newcommand{\he}{\hat{e}}

\newcommand{\hesp}{\hat{e}^{\prime *}}
\newcommand{\hr}{\hat{r}}
\newskip\humongous \humongous=0pt plus 1000pt minus 1000pt
\def\caja{\mathsurround=0pt}

\newif\ifdtup
\def\panorama{\global\dtuptrue \openup1\jot \caja
        \everycr{\noalign{\ifdtup \global\dtupfalse
        \vskip-\lineskiplimit \vskip\normallineskiplimit
        \else \penalty\interdisplaylinepenalty \fi}}}
\def\eqalignno#1{\panorama \tabskip=\humongous
        \halign to\displaywidth{\hfil$\displaystyle{##}$
        \tabskip=0pt&$\displaystyle{{}##}$\hfil
        \tabskip=\humongous&\llap{$##$}\tabskip=0pt
        \crcr#1\crcr}}
\begin{document}
\vspace*{-0.6in}
\hfill \fbox{\parbox[t]{1.12in}{LA-UR-04-8273}}\hspace*{0.35in}
\vspace*{0.0in}

\begin{center}

{\Large {\bf Deuteron Dipole Polarizabilities and Sum Rules}}\\
\vspace*{0.4in}
{\bf J.\ L.\ Friar} \\
{\it Theoretical Division,
Los Alamos National Laboratory \\
Los Alamos, NM  87545} \\
\vspace*{0.20in}
and \\
\vspace*{0.10in}
{\bf G.\ L.\ Payne}\\
{\it Dept. of Physics and Astronomy\\
Univ. of Iowa\\
Iowa City, IA 52242}
\end{center}

\begin{abstract}
The scalar, vector, and tensor components of the (generalized) deuteron electric
polarizability are calculated, as well as their logarithmic modifications.
Several of these quantities arise in the treatment of the nuclear corrections to
the deuterium Lamb shift and the deuterium hyperfine structure. A variety of
second-generation potential models are used and a (subjective) error is assigned
to the calculations. The zero-range approximation is used to analyze a subset of
the results, and a simple relativistic version of this approximation is
developed.
\end{abstract}

\pagebreak
\section{Introduction}
The spectra of hydrogenic atoms and ions have been measured to such high
precision that nuclear properties play a significant role\cite{review}. This is
especially true of the deuterium atom, which has large nuclear contributions to
both the Lamb shift\cite{polar,isotope} and the S-state hyperfine
structure\cite{hyper}. These large nuclear contributions result from the weak
binding of the deuteron and from the concomitant large size of this simplest of
all nuclei.

The Lamb shift can be defined as the difference between calculated energy level
values and those from predefined reference values, which is typically a Dirac
spectrum modified by reduced-mass effects. This shift results after removing
hyperfine splittings, which is equivalent to having a spinless nucleus. 
Hyperfine structure results from the electron's interaction with the nuclear
spin, and can be linear in that spin (the usual type) or quadratic (quadrupole
hyperfine structure). This typically means that the effective nuclear
interaction with the electron can be a scalar, a vector, or a tensor in
character.

The dominant electron-nucleus interactions (beyond the point-nucleus Coulomb
potential) occur in first-Born approximation in their electromagnetic coupling.
Examples are the scalar (or L=0) nuclear finite-size modification of the Coulomb
potential, which dominates nuclear effects in the Lamb shift, and the tensor (or
L=2) modification that dominates the quadrupole hyperfine structure. The
leading-order (vector) hyperfine structure results from the electron's
interaction with the nuclear magnetic moment.

Sub-leading nuclear effects are generated by two-photon processes, which
necessarily involve a loop integral over a virtual photon momentum. The Lamb
shift integral has a weaker dependence on that momentum (i.e., the process is
``softer'') than the corresponding hyperfine process (which is therefore
``harder''). Because the most important parts of the two-photon nuclear
amplitudes involve sequential electron-nucleus electromagnetic interactions,
numerically important contributions arise from inelastic intermediate nuclear
states, and thus are non-static or polarization phenomena. These mechanisms can
also be scalar, vector, or tensor in type.

The leading-order nuclear contribution to the Lamb shift is proportional to the
mean-square nuclear charge radius, while the subleading order is proportional to
the nuclear electric polarizability and its logarithmic modification. The
polarizability is defined in terms of the nuclear electric-dipole operator,
$\vD$, and the fine-structure constant, $\alpha$, as
$$
\alpha_{E} = \frac{2 \alpha}{3} \sum_{N \neq 0}  
\frac{| \langle N| \vD | 0 \rangle |^{2}}{E_{N} - E_{0}}\, , 
\eqno (1)
$$
and its logarithmic mean-excitation energy\cite{polar}, $\bar{E}$,  by
$$
\log (2 \bar{E}/ m_e)\, \alpha_{E} = \frac{2 \alpha}{3} 
\sum_{N \neq 0} \frac{| \langle N | \vD | 0 \rangle |^{2}}{E_{N} - E_{0}} 
\log \left[2 (E_N - E_0)/m_e \right]\, . \eqno (2)
$$
Both are purely non-static, and involve the (virtual) excitation of
negative-parity intermediate states with energy, $E_N$, from the deuteron ground
state with energy, $E_0$. The precise form of the argument of the logarithm has
been dictated by conventional atomic physics usage.

The dominant part of the sub-leading-order nuclear (vector) hyperfine structure
is determined by ``Low'' moments\cite{hyper}, while smaller subleading-order
contributions involve ordinary dipole excitations of a different type. Adopting
a uniform convention for constant factors that will be apparent later we define
$$
\vsig = - i \frac{2 \alpha}{3} \sum_{N \neq 0}  
\langle 0| \vD | N \rangle \times \langle N| \vD | 0 \rangle \equiv
- i \frac{2 \alpha}{3} \langle 0| \vD \times \vD | 0 \rangle \, , \eqno (3)
$$
and
$$
\log (2 \bar{E} / m_e)\, \vsig = - i \frac{2 \alpha}{3} \sum_{N \neq 0}  
\langle 0| \vD | N \rangle \times \langle N| \vD | 0 \rangle 
\log \left[2 (E_{N} - E_{0})/m_e \right] \, . \eqno (4)
$$
The factor of $i$ is necessary to make the appropriate part of $\vsig$ real. We
also note that in non-relativistic approximation $\vD \times \vD$ vanishes
because all components of $\vD$ commute with each other. This is not true when
relativity is incorporated in the calculation (see Eqn.~(34) of \cite{dhg-nuc},
which demonstrates that $\vD \times \vD$ is imaginary). In that event the
commutator contributes an essential part of the derivation of the
Drell-Hearn-Gerasimov sum rule\cite{DHG} and the Low-Energy Theorem\cite{Low}.
The quantity $\vsig$ is more appropriately described as a sum rule, rather than
as a polarizability.

The leading-order contribution to the quadrupole hyperfine structure is driven
by the quadrupole part of the ordinary Coulomb interaction, whose scale is set
by the nuclear quadrupole moment, $Q$. A sub-leading contribution is determined
by the tensor (L=2) component of the nuclear electric polarizability tensor
$$
\alpha_{E}^{\beta \alpha} = \frac{2 \alpha}{3} \sum_{N \neq 0}  
\frac{\langle 0| D^{\beta} | N \rangle \langle N| D^{\alpha} | 0 \rangle}
{E_{N} - E_{0}}\, , \eqno (5a)
$$
which has been calculated only once long ago using separable potentials and the
RSC potential model\cite{tau}. Those calculations were used in a purely nuclear
(as opposed to molecular) determination\cite{Q} of the deuteron's quadrupole
moment. No estimate of the polarizability effect on the determination of the
deuteron's quadrupole moment from the HD molecular hyperfine structure has ever
been made. There will also be a logarithmic modification of this tensor
polarizability, which is obtained by inserting $\log \left[2 (E_{N} - E_{0})/m_e
\right]$ in the sum over intermediate states.

A system with spin 1 can in general have scalar, vector, and tensor
polarizabilities, and these can be of the ordinary type defined in Eqn.~(5a), or
of a type without the factor of $(E_{N} - E_{0})$ in the denominator, determined
by
$$
D^{\beta \alpha}\,= \frac{2 \alpha}{3} \langle 0 | D^\beta D^\alpha | 0 \rangle
\, , \eqno (5b)
$$
which was illustrated by Eqn.~(3) (and more accurately called a sum rule). Each
of these 6 types can have a logarithmic modification. Because the deuteron is an
important nucleus and deuterium an important atom, we will calculate all twelve
of these polarizabilities (or sum rules) using modern second-generation
potentials. Only three of the polarizabilities have been calculated before, and
only two with modern potentials. None of the logarithmic modifications of the
sum rule above have been previously calculated. Our approach will be
non-relativistic in keeping with the potential models we use. Our calculations
will be based upon Podolsky's method\cite{boris} for computing second-order
perturbation-theory matrix elements, and on several integration tricks, one of
which has been used in the past.

\section{Tensor Polarizabilities}

The electric polarizability is most easily calculated using second-order
perturbation theory and the coupling of the nuclear electric-dipole-moment
operator to a uniform electric field, which leads immediately to Eqn.~(5a). This
equation is fully equivalent to
$$
\alpha_{E}^{\beta \alpha}  = \frac{2 \alpha}{3} \; \langle 0 | D^{\beta} | 
\Delta \Psi^{\alpha} \rangle \, , \eqno (6a)
$$
where\cite{boris}
$$
(H - E_0) \; | \Delta \Psi^{\alpha} \rangle \; = \; D^{\alpha} | 0 \rangle 
\eqno (6b)
$$
is solved subject to finite boundary conditions.  Note that $\vec{D}$ does not
connect the ground state (the only bound state) of the deuteron to itself.
Resolution of Eqns.~(6) into partial waves is necessary in order to perform a
numerical calculation.

Because we wish to compute all tensor components of the electric polarizability
tensor we present a very brief derivation of the partial-wave decomposition that
updates older work\cite{ae,polar}. We begin with Eqns.~(5a) and (6) and convert
them to a scalar by contracting the Cartesian index $\alpha$ with a constant
vector $E^{\alpha}$ and the index $\beta$ with a constant vector $E^{\prime \,
\beta}$. The deuteron initial state $\Psi_d = |S M \rangle$ (and analogously for
the final state) depends on the azimuthal quantum number $M$ and we formally
remove it from the problem by defining\cite{edmonds} a (vector) projection
operator (appropriate for $S=1$): $e^{*}_{1 \lambda} (M) \equiv 
\delta_{\lambda , M}$. Performing the sum $\sum_\lambda e^{*}_{ 1 \lambda} (M)
\, |S \lambda \rangle$ results formally in a scalar quantity independent of spin
projections and amenable to manipulation. A similar projector
$\sum_{\lambda^\prime} e^{*}_{1 \lambda^\prime} (M^\prime)\, \langle S
\lambda^\prime |$ is used for the final (i.e., leftmost) deuteron state.

We define general orthonormal spin-angular wave functions for the deuteron 
system
$$
\phi^{\ell}_{J M} = (Y_{\ell} (\hr) \otimes \chi_1)_{J M} \, , \eqno (7)
$$
which couples the usual angular wave function that depends on the direction
$\hr$ of the internucleon vector $\vr$ to the (unit-) spin wave function,
$\chi$. The deuteron's full wave function is then given by
$$
\psi_d = \sum_{\ell = 0,2} a_{\ell} (r)\, \phi^{\ell}_1 (\hr) \, , \eqno (8)
$$
where $a_0 (r) = u(r)/r$ and $a_2 (r) = w(r)/r$ expresses $a_\ell$ in terms of
the conventional radial wave function components of the deuteron. We have
suppressed for now the azimuthal quantum number, $M$.

We ignore the tiny effect of the n-p mass difference on the deuteron
center-of-mass (CM), and find $\vD = \forth (\tau_1^z - \tau_2^z)\, \vr$
expressed in terms of the isospin operators $\btau_i$ of nucleon $i$. The dipole
isospin operator generates $T=1$ states when acting on the deuteron, together
with a residual numerical factor of $\haf$ that we will ignore until later. If
the right-hand-side (RHS) of Eqn.~(6b) (contracted with $E^\alpha$ and $e^{*}_{1
\lambda} (M)$) is expanded in terms of $\phi_J^{\ell}$ functions, one finds
$$
{\rm RHS} = \sum_{\stackrel{\scriptstyle L = 1,3}{J=0,1,2}} \phi_J^L \cdot 
\left( E_1 \otimes \he_1 \right)_J \,(-1)^J \, \frac{g^L_J (r)}{\sqrt{3}} 
\, , \eqno (9)
$$
where
$$\eqalignno{
g^1_0 &= u-  \sqrt{2}\, w &(10a) \cr
g^1_1 &= u+    w/\sqrt{2} &(10b) \cr
g^1_2 &= u-\sqrt{2}\,w/10 &(10c) \cr
g^3_2 &=  3 \sqrt{3}\, w/5&(10d) \cr}
$$
are the relevant radial functions. We manipulate the left-hand-side of Eqn.~(6b)
(similarly contracted) into the same form as Eqn.~(9)
$$
|\Delta \Psi^\alpha \rangle E^\alpha = \frac{- 2 \mu r}{r} \sum_{\stackrel{
\scriptstyle L = 1,3}{J=0,1,2}} \phi_J^L \cdot \left( E_1 \otimes \he_1
\right)_J \,(-1)^J \, \frac{f^L_J (r)}{\sqrt{3}} \, , \eqno (11)
$$
where $\mu$ is the n-p reduced mass and the functions $f^L_J (r)$ satisfy
$$
( H_{L J} - E_d )\frac{-2 \mu r}{3} f^L_J (r) = g^L_J   \, . \eqno (12)
$$ 
We note that for total angular momentum $J=2$ the $L=1$ and $L=3$ orbital
components are coupled by the tensor force. We do treat that coupling properly,
although it is not reflected in the simplified notation employed in Eqn.~(12).

The matrix element in Eqn.~(6a) (including its factor of ($2 \alpha/3$), a
factor of $\haf$ from each of the two dipole operators, the factors of ($-2
\mu$) and 1/$\sqrt{3}$ from Eqn.~(11), and the $1/\sqrt{3}$ from Eqn.~(9)) then
becomes
$$
\alpha_E^{\beta \alpha} \rightarrow \sum_{J} {(E_1^\prime \otimes 
\hesp_1 )}_J \cdot {(E_1 \otimes \he_1 )}_J \, (-1)^J \, \sum_L a^L_J \, , 
\eqno (13a)
$$
where
$$
a^L_J = \frac{- \mu \,\alpha}{9} \int_0^\infty d r\, r^2 f^L_J (r)\, g^L_J (r)
\eqno (13b)
$$
now expresses the entire content of the electric polarizability tensor in terms
of projection operators and matrix elements. Equation~(13a) is not a convenient
form, and we recouple it so that the projectors of the same type are coupled
together. We also note that only the $J=2$ part of the $a^L_J$ terms has two
non-vanishing components. We therefore define
$$
A_J = a^1_J + \delta_{J,2}\, a^3_2 \, , \eqno (14)
$$
and
$$
\lambda_\kappa = \left( E_1^\prime \otimes E_1 \right)_ \kappa \cdot 
\left( \hesp_1 \otimes \he_1 \right)_\kappa \, , \eqno (15)
$$
which yields
$$
\alpha_{E}^{\beta \alpha}  \rightarrow \sum_{\kappa = 0,1,2} \lambda_\kappa
b_{\kappa} \, , \eqno (16)
$$
where the (real) quantities
$$\eqalignno{
b_0 &= \frac{1}{3} (A_0 + 3 A_1 +5 A_2)    &(17a) \cr
b_1 &= \frac{1}{6} (-2 A_0 - 3 A_1 +5 A_2) &(17b) \cr
b_2 &= \frac{1}{6} (2 A_0 - 3 A_1 + A_2)   &(17c) \cr}
$$
determine the tensor properties of the nuclear physics.

The structure of the $\lambda_\kappa$ operators corresponds to tensors of order
$\kappa$ in both the Cartesian indexes $\alpha$ and $\beta$, and in the
effective (azimuthal) spin dependence. The dependence on the two spin-projection
operators $\he_1$ and $\hesp_1$ is indeed equivalent to using the Wigner-Eckart
Theorem\cite{edmonds} on the nuclear matrix elements. This allows us to rewrite
the spin factors $\lambda_\kappa$ as effective operators in the nucleus
total-angular-momentum Hilbert space, determined by powers of the
angular-momentum operator, $\vS$ of the nucleus:
$$
\alpha_{E}^{\beta \alpha} = \alpha_E \frac{\delta^{\alpha \beta}}{3}
+ i \sigma \epsilon^{\beta \alpha \gamma} \frac{S^\gamma}{2}
+ \tau \left( \frac{S^\alpha S^\beta + S^\beta S^\alpha}{2} - 
\frac{2 \delta^{\alpha \beta}}{3} \right) \, . \eqno (18)
$$
The three coefficients $\alpha_E$, $\sigma$, and $\tau$ are, respectively, the
scalar, vector, and tensor components of the polarizability. Equation (1) (a
trace) defines $\alpha_E$, while $\sigma$, and $\tau$ are defined by
$$
\sigma = - i \frac{2 \alpha}{3} \; \epsilon^{\lambda \mu 3} \sum_{N \neq 0}  
\frac{\langle S S| D^\lambda | N \rangle \langle N| D^\mu 
| S S \rangle} {E_{N} - E_{0}}\, , \eqno (19a)
$$
and
$$
\tau = 3 \alpha_E^{3 3} -\alpha_E \, , \eqno (19b)
$$
where the deuteron should be in the state $M^{\prime}= M = S = 1$. Finally the 
relationships
$$\eqalignno{
\alpha_E =& b_0  &(20a)\cr
\sigma   =& -b_1  &(20b)\cr
\tau     =& -b_2 &(20c)\cr}
$$
determine the various polarizabilities in terms of the partial waves. Note that
$\sigma$ is real, since the Cartesian vector cross product differs from the
spherical one by a factor of $-i \sqrt{2}$ and the Wigner-Eckart Theorem
guarantees that the spherical result is overall real.

The sum-rule quantity $D^{\beta \alpha}$ is decomposed in strict analogy to
Eqn.~(18), with coefficients, $s$, $v$, and $t$ replacing $\alpha_E$, $\sigma$,
and $\tau$. In the non-relativistic approximation that we employ (or,
equivalently in the deuteron, the impulse approximation) the former quantities
are related to conventional deuteron moments by $s = \frac{2 \alpha}{3} \langle
r^2 \rangle_{ch}$ $ v = 0$, and $t = \frac{2 \alpha}{3} Q$, where $\langle r^2
\rangle_{ch}$ is the mean-square charge radius and $Q$ is the quadrupole moment.
Note that in the representation of Eqn.~(18) the quantity $\vsig$ in Eqn.~(3)
becomes $\vsig = v\, \vS$.

\section{Logarithmic Sum Rules}

Calculations of the logarithmic modification of the basic polarizabilities or
sum rules use the trick of adding an arbitrary energy $\xi f$ to the energy
denominator in Eqn.~(5a), where $f$ has the dimensions of energy and $\xi$ is
dimensionless. This defines
$$
\alpha_{\rm E}^{\beta \alpha} (\xi ) = \frac{2 \alpha}{3} \sum_{N \neq 0}  
\frac{\langle 0| D^{\beta} | N \rangle \langle N| D^{\alpha} | 0 \rangle}
{\xi f + E_{N} - E_{0}}\, . \eqno (21)
$$
We first integrate $\alpha_{E}^{\beta \alpha} (\xi ) $ with respect to $\xi$
from 0 to $\Lambda$, where $\Lambda$ is very large compared to any relevant 
energies $(E_N - E_0)$. This produces
$$
f \int_0^\Lambda d \xi \, \alpha_{E}^{\beta \alpha} (\xi ) = \frac{2 
\alpha }{3} \sum_{N \neq 0} 
\langle 0| D^{\beta} | N \rangle \langle N| D^{\alpha} | 0 \rangle
\log{[\Lambda f/(E_{N} - E_{0})]} \, . \eqno (22)
$$
We split the integration region from $[0, \Lambda]$ into $[0, 1]$ plus $[1,
\Lambda]$. In the second region we change variables from $\xi$ to $1/\xi$. We
also note that $\int^1_{1/\Lambda} d \xi/\xi = \log{\Lambda}$. Putting
everything together we find that $\Lambda$ can be taken to infinity if we use
$$\eqalignno{
\frac{2 \alpha}{3} \sum_{N \neq 0} &
\langle 0| D^{\beta} | N \rangle \langle N| D^{\alpha} | 0 \rangle
\log{[2 (E_{N} - E_{0})/m_e]} \equiv D^{\beta \alpha} \log{(2 \bar{E}/m_e)}=
&{}\cr
-& \int_0^1 d \xi \, f\, \alpha_{E}^{\beta \alpha} (\xi ) 
- \int_0^1 \frac{d \xi}{\xi}\, \left[ 
\frac{f}{\xi}\, \alpha_{E}^{\beta \alpha} (1/\xi ) - D^{\beta \alpha} 
\right] +D^{\beta \alpha}\, \log{(2 f/m_e)} \, ,  &(23)\cr}
$$
where $D^{\beta \alpha}$ is defined in Eqn.~(5b).

A similar set of manipulations was developed previously in which the integral
$$
\int_\epsilon^\infty \frac{d \xi}{\xi}\, \alpha_{E}^{\beta \alpha} (\xi ) 
= \frac{2 \alpha}{3} \sum_{N \neq 0}
\frac{\langle 0| D^{\beta} | N \rangle \langle N| D^{\alpha} | 0 \rangle}
{E_{N} - E_{0}} \log{[(E_{N} - E_{0})/\epsilon f]}  \eqno (24)
$$
was split into the regions [$\epsilon$,1] plus [1,$\infty$], and the integration
variable for the second region was also changed to 1/$\xi$. This led to a
special case of
$$
\alpha_E^{\beta \alpha} (0) \, \log \; (2 \bar{E}/m_e) = \int^{1}_{0} \; 
\frac{d \xi}{\xi} [ \alpha_E^{\beta \alpha} (\xi) - \alpha_E^{\beta \alpha}
(0) + \alpha_E ^{\beta \alpha} ( 1 / \xi )] - \alpha_E^{\beta \alpha} (0)
\, \log (m_e/2 f) \, . \eqno (25)
$$
Equations (23) and (25) provide a tractable scheme for calculating logarithmic
modifications of our basic polarizability, $\alpha_E^{\beta \alpha} (0)$, and of
the sum-rule quantity $D^{\beta \alpha}$. The results are independent of the
scale parameter, $f$. Note also that since the various polarizabilities are
particular linear combinations of partial waves, the logarithmic modifications
are also the same linear combinations involving those partial waves, and this is
a convenient way to perform the calculations. We will also see in the next
section that the zero-range approximation provides an excellent starting point
for understanding the scalar polarizabilities (the vector and tensor ones are
significantly smaller and vanish in this approximation). In the Appendix we
develop the zero-range form of Eqn.~(5a), which allows the analytic calculation
of the logarithmic modifications to $\alpha_E$ and $D^{\alpha \alpha}$. We also
develop a zero-range model based on the Relativistic Schr\"odinger Equation
(RSE).

\section{Results and Conclusions}

We have calculated the scalar, vector, and tensor components of both
$\alpha_E^{\beta \alpha}$ and $D^{\beta \alpha}$, together with their
logarithmic modifications. These calculations were performed with seven
different second-generation potential models, including the Argonne V$_{18}$
(AV18) \cite{2gen2}, the Reid Soft Core 1993 (RSC93), and 5 Nijmegen
models\cite{2gen0,2gen1}, including the full model (no partial-wave expansion)
as well as the local and non-local Reid-like models. The last two types had been
fitted to both relativistic and non-relativistic forms of the deuteron binding
energy.

It has been known for a long time that the deuteron mean-square radius and the
electric polarizability are rather accurately predicted by the zero-range
approximation (see the Appendix). That approximation over-predicts the
polarizability by approximately 1\% and $D^{\alpha \alpha}$ by less than 2\%,
and is therefore an excellent starting point for investigating the uncertainties
in the four scalar quantities. The largest uncertainty in the zero-range results
is due to $A_S$, the asymptotic S-wave normalization constant, whose value was
determined in phase-shift analyses\cite{AS} to be $A_S$ = 0.8845(8) fm$^{-1/2}$.
This leads via Eqn.~(A6) to the zero-range result, $\alpha_E^{\rm zr}$ =
0.6378(12) fm$^{3}$, and via Eqn.~(\A2) to $D^{\alpha \alpha}_{\rm zr} =
0.01916(4)$ fm$^2$ , where we use the relativistic form of the deuteron binding
energy for both quantities.

\begin{table}[htb]
\centering
{\bf Table I}
\caption{Scalar, vector, and tensor components of $D^{\beta \alpha}$ (viz., $s$,
$v$, $t$) and $\alpha_E^{\beta \alpha}$ (viz., $\alpha_E$, $\sigma$, $\tau$),
followed by the product with the appropriate logarithmic factor, $\log{(2
\bar{E}/m_e)}$, for each case. Results were calculated using a number of
different potential models as discussed in the text, and the ``error bar''
results from combining a subjective estimate of the spread in the results after
scaling to the experimental values of $A_S^2$ and $\eta$, as discussed in the
text, with the variances of those two quantities. All calculations used the
impulse approximation for the dipole operator and assumed equal-mass nucleons.
Note that the first two rows have been multiplied by a factor of 10 to make the
entries more uniform in size.}
\vspace*{0.1in}
\begin{tabular}{| c || c c c l|}
\hline
type        & Scalar & Vector & Tensor~ &{} \\ \hline
$D^{\beta \alpha} \times 10$  
  & 0.1882(4)~ & 0.00000~ &~ 0.01322(10) & fm$^2$ 
   \rule{0in}{2.5ex}\\ \hline 
$D^{\beta \alpha} \log{(2 \bar{E}/m_e)} \times 10$ 
  & 0.6327(12)& ~0.0003(1) & 0.0503(4)~~ & fm$^2$ 
   \rule{0in}{2.5ex}\\ \hline
$\alpha_E^{\beta \alpha}$         
  & 0.6330(13) &\minus 0.00092(5)~& 0.0317(3) ~ & fm$^3$ 
   \rule{0in}{2.5ex}\\ \hline
$\alpha_E^{\beta \alpha} \log{(2 \bar{E}/m_e)}$ 
  & 1.8750(36) & \minus 0.0023(2)~ & 0.1014(8) ~ & fm$^3$
   \rule{0in}{2.5ex}\\ \hline
\end{tabular}
\end{table}

The second-generation potentials are sufficiently accurate that they can be
regarded as alternative phase-shift analyses. We therefore expect that the
values of the electric polarizability and $D^{\alpha \alpha}$ will scatter
around a central value with the variance of 2 parts/thousand associated with
$A_S^2$. In order to check whether the potential-model variation of the
remaining 1\% of $\alpha_E$ and roughly 2\% of $D^{\alpha \alpha}$ is small, we
have scaled each calculated quantity by $(A_S^{\rm exp}/A_S^{\rm model})^2$, and
have examined the remaining variations. The scaled values are listed in Table I.
This procedure verifies that the dominant uncertainty in the scalar quantities
is the error in $A_S$.

The tensor quantities should be expected to scale like $\eta$, the asymptotic
$D/S$ (amplitude) ratio of the deuteron. Those quantities in Table I have
therefore also been scaled by $\eta^{\rm exp}/\eta^{\rm model}$, where the
experimental value of $\eta$ was determined from phase-shift analyses\cite{AS}
to be $\eta^{\rm exp} = 0.0253(2)$. This scaling considerably reduces the
scatter in the calculated tensor results. The small size of $\eta$ also roughly
accounts for the size of the tensor quantities relative to the scalar ones. The
result in Table~I is nearly 10\% smaller than the one in Ref.~\cite{tau},
reflecting a smaller modern value of $\eta$ \cite{AS}.

The vector quantities are very suppressed and are sensitive to details of the
nuclear force that vary from model to model. They are sensitive to the forces in
the $^3P_J$ states, and in the absence of those forces can be shown to be
determined by the square of the D-state wave function. The quoted uncertainties
are inferred solely from the model variations. Note that the vector part of
$D^{\beta \alpha}$ vanishes identically in non-relativistic approximation, and
this accounts for the very small sizes of the vector quantities.

In summary, we have calculated a variety of polarizabilities and sum rules for
the deuteron that are generated by (unretarded) electric dipole interactions.
These quantities have been divided into scalar, vector, and tensor components,
and include logarithmic modifications of each. Our numerical techniques allow us
to generate all such components at no extra cost. The scalar polarizabilities
were previously calculated\cite{polar}, and have not changed significantly. The
vector components of the sum rule play a role in the ordinary deuterium
hyperfine structure\cite{hyper} and are very small. The tensor components
determine part of the deuterium quadrupole hyperfine structure\cite{review},
although calculations of this effect have not yet been performed. The sizes and
uncertainties of the various quantities were analyzed using the zero-range
approximation and various scales appropriate to the deuteron.

\section{Appendix}

The zero-range approximation is motivated by the asymptotic dominance of radial
matrix elements that contain (positive) powers of the distance between the
nucleons. The mean-square radius is an obvious example, as is the electric
polarizability, since each is weighted by two powers of the inter-nucleon
separation. The usual version of this approximation is to assume that in
intermediate states the nucleons lie outside the range of the nuclear force
(i.e., we set that force to zero) and in the initial and final deuteron states
we ignore the D-wave and use the asymptotic form of the S-wave function. The
zero-range deuteron wave functions are therefore given by
$$\eqalignno{
u_{\rm zr} (r) =& A_S \exp{(-\kappa r)}\, , &(\A1a) \cr
w_{\rm zr} (r) =& 0\, ,                     &(\A1b) \cr}
$$
where $A_S$ is the asymptotic S-wave normalization constant, $\kappa = \sqrt{2
\mu E_b}$, $\mu$ is the n-p reduced mass, and $E_b$ is the (positive) deuteron
binding energy. Ignoring the tiny difference in the proton and neutron masses
(the neutron lies slightly closer to the CM than the proton) one finds that the
mean-square charge radius and $\langle \vD^2 \rangle$ are proportional in 
impulse approximation (which we have assumed as a consequence of the
non-relativistic approximation), but not otherwise. Recalling the factor of $(2
\alpha/3)$ from Eqn.~(5b) and two factors of $\haf$ from the two dipole
operators, the zero-range value of $D^{\alpha \alpha}$ is given by
$$
D^{\alpha \alpha}_{\rm zr} = \frac{\alpha A_S^2}{24\, \kappa^3} \rightarrow
0.01916(4)\, {\rm fm^2} \, , \eqno (\A2)
$$
where the numerical result uses the experimental value of $A_S$ and the
relativistic value of $\kappa$ defined below. The calculation of the remaining
quantities requires an analytic expression for the scalar electric
polarizability with the $\xi f$ insertion. It is convenient to choose $f$ to be
the deuteron binding energy (i.e., $E_b = - E_0$), so that in momentum space we
have
$$
\xi E_b +E_N - E_0 \rightarrow \frac{(1 + \xi)\, \kappa^2 + \vp^2}{2 \mu} 
\, , \eqno (\A3)
$$
and the required Green's function is therefore a simple modification of the 
usual zero-range Green's function for the deuteron:
$$\eqalignno{
G_0 (\xi) =& \frac{2 \mu}{4 \pi r} \exp{(- \bar{\kappa}\, r)} &(\A4a)\cr
\bar{\kappa} =& \sqrt{1 + \xi}\, \kappa \, .                  &(\A4b)\cr}
$$
Performing the integrals in Eqn.~(5a) using Eqn.~(\A1) for the wave function and
Eqn.~(\A4) for the Green's function leads to the zero-range result
$$
\alpha_E^{\rm zr} (\xi) = \frac{\alpha \mu A_S^2}{12 \kappa^3} 
\frac{\left( \kappa^2 + \bar{\kappa}^2 + 4 \kappa \bar{\kappa}\right)}
{(\kappa + \bar{\kappa})^4} \, . \eqno (\A5)
$$
This gives the well-known result 
$$
\alpha_E^{\rm zr} (0) = \frac{\alpha \mu A_S^2}{32 \kappa^5} \rightarrow
0.6378(12)\, {\rm fm^3} \, , \eqno (\A6)
$$
where the numerical result uses the experimental value of $A_S$ and the
relativistic value of $\kappa$ defined below. Equations (22) and (24) can now be
used to calculate the logarithmic modifications of $\alpha_E^{\rm zr}$ and
$\langle \vD^2 \rangle_{\rm zr}$. The logarithmic modification of $\alpha_E^{\rm
zr}$ is determined by
$$
\log{(\bar{E}/E_b)} = \log{4} -\frac{7}{12} \, , \eqno (\A7a)
$$
or $ \frac{\bar{E}}{E_b} = 2.23214 \cdots$, while the modification of $\langle
\vD^2 \rangle_{\rm zr}$ is determined by
$$
\log{(\bar{E}/E_b)} = \log{4} -\frac{1}{6} \, , \eqno (\A7b)
$$
or $ \frac{\bar{E}}{E_b} = 3.38592 \cdots$. Both results are very simple and
quite accurate.

Our final task is to estimate the size of one class of relativistic corrections
to $\alpha_E^{\rm zr}$. We begin with the so-called Relativistic Schr\"odinger
Equation, which we construct for two non-interacting nucleons with identical
masses, $M$, by summing the kinetic energies of each:
$$
\left[E-\left(M^2 + \vp^2_1 \right)^{\haf} - \left( M^2 + \vp^2_2 \right)^{\haf}
\right] \Psi = 0 \, . \eqno (\A8)
$$
A potential could also be added to the kinetic energy. In the center-of-mass
frame of the two particles (with momenta $\vp$ and $-\vp$, respectively)
$$
\left[E_{cm} - 2 \left(M^2 + \vp^2 \right)^{\haf} \right] \Psi = 0 \, , 
\eqno (\A9a)
$$
indicating that the energy of a bound deuteron would be given by
$$
E_d = 2 \left(M^2 - \kappa_r^2 \right)^{\haf} \equiv 2 M - E_b \, . \eqno (\A9b)
$$
Since the rightmost (experimental) result holds in all cases, it clearly makes a
difference if the non-relativistic approximation $2 M - \kappa_{nr}^2/M$ is
substituted for the square root in Eqn.~(\A9b). For this reason we have labelled
the relativistic value of $\kappa$ as $\kappa_r$ and the non-relativistic
approximation as $\kappa_{nr}$. They are related by $\kappa_r \cong \kappa_{nr}
(1-\kappa^2_{nr}/8M^2)$.

Equation~(\A9a) also holds if we multiply it by $E_{cm} + E_{cm}^{\prime}$,
where $E_{cm}^{\prime} = 2 \left(M^2 + \vp^2 \right)^{\haf}$. This reduces that
equation to non-relativistic {\it form}, but with $\kappa_r$ replacing
$\kappa_{nr}$. Equations (\A1) therefore still hold {\it mutatis mutandis}. This
does not apply to the Green's function, however. If we invert Eqn.~(\A9a) and
multiply top and bottom by the identical factor $E_{cm} + E_{cm}^{\prime}$, the
denominator has the desired form $- 4(\kappa^2_r +\vp^2)$, but the numerator now
contains the factor $(E_{cm} + E_{cm}^{\prime})$, which we rewrite as $2E_{cm}
+(E_{cm}^{\prime}-E_{cm})$ and note that the second part of this expression
cancels a similar factor in the denominator. The remainder is very short ranged
(range $\sim 1/M$) when Fourier transformed, and in keeping with the zero-range
approximation we ignore this term. Thus the appropriate zero-range Green's
function for the RSE is simply $\left( 1 - \kappa^2_r/M^2 \right)^{1/2} G_0$,
where the form in Eqn.~(\A4a) holds if we replace $2 \mu$ by $M$ and $\kappa$ by
$\kappa_r$. Thus previous results for $\alpha_E$ hold is we use $\kappa_r$ 
everywhere for $\kappa$ and multiply by the factor of $\left( 1 - \kappa^2_r/M^2
\right)^{1/2}$. This produces a correction (compared to 1) $\sim - \kappa_r^2/2
M^2 \sim -0.0012$, which reflects the expected size of relativistic effects for
the deuteron. For a full treatment, see Ref.\cite{chi-pt}.

\section*{Acknowledgments}

The work of JLF was performed under the auspices of the U.\ S.\ Dept.\ of
Energy, while the work of GLP was supported in part by the DOE.

\end{document}